\documentclass[prd,nofootinbib,aps,floats,floatfix,amsmath,amssymb,secnumarabic,onecolumn]{revtex4} %

\usepackage{graphicx}
\usepackage[usenames,dvipsnames]{color}
\usepackage{amsmath,amssymb}
\usepackage{slashed}
\usepackage{verbatim}
\usepackage[colorlinks,citecolor=blue]{hyperref}
\allowdisplaybreaks[1]

\begin{document}

\title{\boldmath New Physics effect on $B_c \to J/\psi \tau\bar\nu$ in relation to the $R_{D^{(*)}}$ anomaly }

\author{Ryoutaro Watanabe}
\email{watanabe@lps.umontreal.ca}
\affiliation{Physique des Particules, Universit\'e de Montr\'eal, \\ C.P. 6128, succ.\ centre-ville, Montr\'eal, QC, Canada H3C 3J7}

\begin{abstract} 
We study possible new physics (NP) effects on $B_c \to J/\psi \tau\bar\nu$, which has been recently measured at LHCb as the ratio of $R_{J/\psi} = \mathcal B(B_c \to J/\psi \tau\bar\nu)/\mathcal B(B_c \to J/\psi \mu\bar\nu)$. 
Combining it with the long-standing $R_{D^{(*)}}$ measurements, in which the discrepancy with the prediction of the standard model is present, we find possible solutions to the anomaly by several NP types. 
Then, we see that adding the $R_{J/\psi}$ measurement does not improve NP fit to data, but the NP scenarios still give better $\chi^2$ than the SM. 
We also investigate indirect NP constraints from the lifetime of $B_c$ and NP predictions on the $\tau$ longitudinal polarization in $\bar B \to D^* \tau\bar\nu$. 
{\tt ~~~~~~~~~~~~~~~~~~~~~~[UdeM-GPP-TH-17-259]}
\end{abstract}
\maketitle

\section{Introduction}
\label{Intro}

On recent years, discrepancies with the predictions of the Standard Model (SM) have started to emerge in semi-tauonic decays of $B$ meson, $\bar B \to D^{(*)}\tau\bar\nu$. 
Measurements have been done in the lepton-universality ratios, 
\begin{align}
 R_{D^{(*)}} = \frac{\mathcal B(\bar B \to D^{(*)} \tau\bar\nu)}{\mathcal B(\bar B \to D^{(*)} \ell\bar\nu)} \,, 
\end{align}
for $\ell = e$ or $\mu$. 
The world average of the BaBar~\cite{Lees:2012xj,Lees:2013uzd}, Belle~\cite{Huschle:2015rga,Sato:2016svk,Abdesselam:2016xqt}, and LHCb~\cite{Aaij:2015yra,Aaij:2017uff} results 
shows $\sim 4\sigma$ deviation from the SM prediction. 
Then, many theorists have tried to address this anomaly in different new physics (NP) models; 
as in Refs.~\cite{Sakaki:2013bfa,Sakaki:2014sea,Tanaka:2012nw,Fajfer:2012vx,Becirevic:2012jf,Datta:2012qk,Duraisamy:2013kcw,Biancofiore:2013ki,Bhattacharya:2015ida,Alok:2016qyh,Feruglio:2016gvd,Faroughy:2016osc,Bardhan:2016uhr,Ivanov:2017mrj,Ivanov:2016qtw,Choudhury:2017qyt} for model-independent approaches, Refs.~\cite{Hou:1992sy,Tanaka:1994ay,Kiers:1997zt,Chen:2006nua,Crivellin:2012ye,Celis:2012dk,Crivellin:2013wna,Cline:2015lqp,Kim:2015zla,Crivellin:2015hha,Wang:2016ggf,Celis:2016azn,Iguro:2017ysu} for charged Higgs, Refs.~\cite{Ko:2012sv,Alonso:2015sja,Freytsis:2015qca,Barbieri:2015yvd} for lepton flavor violation, Refs.~\cite{Sakaki:2013bfa,Calibbi:2015kma,Fajfer:2015ycq,Bauer:2015knc,Hati:2015awg,Sahoo:2015pzk,Zhu:2016xdg,Dumont:2016xpj,Das:2016vkr,Li:2016vvp,Bhattacharya:2016mcc,Becirevic:2016yqi,Sahoo:2016pet,Bhattacharya:2014wla} for leptoquarks (in relation to $B\to K^{(*)}\mu^+\mu^-$), and Refs.~\cite{Boucenna:2016wpr,Boucenna:2016qad,Altmannshofer:2017poe} for others.  
%
When we start with the low-energy effective field theory, NP effects are described by the four fermion operators of $(bc\tau\nu)$: 
\begin{align}
 -\mathcal L = {4G_F \over \sqrt2} V_{cb} 
 & \left[ (1 + C_{V_1}) (\bar c_L \gamma^\mu b_L)(\bar\tau_L \gamma_\mu \nu_{L}) + C_{V_2}(\bar c_R \gamma^\mu b_R)(\bar\tau_L \gamma_\mu \nu_{L}) \right. \notag \\
 & \left. + C_{S_1} (\bar c_L b_R)(\bar\tau_R \nu_{L}) + C_{S_2} (\bar c_R b_L)(\bar\tau_R \nu_{L}) + C_T (\bar c_R \sigma^{\mu\nu} b_L)(\bar\tau_R \sigma_{\mu\nu} \nu_{L}) \right] \,, \label{eq:Leff}
\end{align}
where NP effects are encoded in the Wilson coefficients $C_X$.  
At the present stage, NP contributions with nonzero $C_X$ from single operators in \eqref{eq:Leff} are possible solutions to the $R_{D^{(*)}}$ anomaly except for $C_{S_1}$\footnote{
The $S_1$ type operator $(\bar c_L b_R)(\bar\tau_R \nu_{L})$ never accommodates the experimental values of $R_{D}$ and $R_{D^*}$ at the same time. 
Henceforth, we skip the $S_1$ scenario in this paper from the beginning. 
}, according to the previous studies, {\it e.g.}, as in Refs.~\cite{Sakaki:2013bfa,Sakaki:2014sea}. 
The $V_1$ scenario has an advantage such that a similar $V-A$ current in the $bs$ system can also explain the anomalies in $B\to K^{(*)}\mu^+\mu^-$, 
({\it e.g.}, see Refs.~\cite{Bhattacharya:2014wla,Calibbi:2015kma,Bhattacharya:2016mcc}.) 
The $V_2$ scenario requires $C_{V_2}$ to be pure imaginary and the $S_2$ scenario needs a large negative $C_{S_2}$, to address the $R_{D^{(*)}}$ anomaly~\cite{Tanaka:2012nw,Crivellin:2012ye}.

Some leptoquark (LQ) models contribute to $\bar B \to D^{(*)} \tau\bar\nu$ with scalar-tensor operators so that $C_{S_2} \simeq \pm 7.8 C_T$ at the $m_b$ scale\footnote{
At the scale where LQ models are defined, the corresponding relations are $C_{S_2} = \pm 4 C_T$. 
These relations are realized for the scalar leptoquark bosons ($R_2$ and $S_1$) that transform as $({\bf 3}, {\bf 2}, 7/6)$ and $(\bar{\bf 3}, {\bf 1}, 1/3)$ under $SU(3)_c \times SU(2)_L \times U(1)_Y$, respectively. 
}.
Then, they also explain the $R_{D^{(*)}}$ anomaly. 
For a dedicated study, see Ref.~\cite{Sakaki:2013bfa}.

In Ref.~\cite{Alonso:2016oyd}, this anomaly has been investigated by looking at the lifetime of $B_c$ meson along with the decay $B_c \to\tau\bar\nu$. 
As $C_{X} \neq 0$ (for $X \neq T$) also contributes to $B_c \to\tau\bar\nu$, it is necessary that the contribution does not exceed the fraction of the total decay width of $B_c$, 
which has been experimentally measured and theoretically calculated. 
Indeed, this could allow us to exclude a large contribution from $C_{S_{i}}\neq 0$. 
In Ref.~\cite{Akeroyd:2017mhr}, a stronger limit on the scalar contribution has been suggested with using LEP1 data for $B_c \to\tau\bar\nu$.

In September 2017, the LHCb collaboration reported a new measurement regarding $b \to c\tau\nu$ in $B_c$. 
To be specific, the ratio 
\begin{align}
 R_{J/\psi} 
 = \frac{\mathcal B(B_c \to J/\psi \tau\bar\nu)}{\mathcal B(B_c \to J/\psi \mu\bar\nu)} 
 = 0.71 \pm 0.17 \pm 0.18 \,, 
 \label{eq:RJpsiLHCb}
\end{align}
has been obtained with dataset of run 1 ($3\,\text{fb}^{-1}$)~\cite{LHCbRJpsi,LHCbStatus}. 
Thus, this new measurement enables us to develop explanations for the anomaly with the above NP scenarios, which will be shown in this paper. 
We will also revisit the constraints with use of the lifetime of $B_c$ and put some predictions on the $\tau$ longitudinal polarization.

This letter is then organized as follows. 
In Sec.\,\ref{Sec:formula}, we obtain a formula for the decay rate of $B_c \to J/\psi \tau\bar\nu$ in the presence of the NP operators. 
A description of form factors for the $B_c \to J/\psi$ transition is also given. 
In Sec.\,\ref{Sec:analysis}, we proceed to numerical analysis and obtain possible solutions to the $R_{D}$, $R_{D^*}$, and $R_{J/\psi}$ measurements by the NP scenarios. 
We also investigate NP effect on the lifetime of $B_c$, associated with $B_c \to\tau\bar\nu$, and the $\tau$ longitudinal polarization in $\bar B \to D^* \tau\bar\nu$. 
The Sec.\,\ref{Sec:summary} is devoted to summary.

\section{Description of hadronic amplitude and form factors}
\label{Sec:formula}

The hadronic transition of $B_c \to J/\psi$ can be written in analogy with that of $\bar B \to D^{*}$. 
Namely, we can obtain the formula for the decay rate of $B_c \to J/\psi \tau\bar\nu$ as follows~\cite{Sakaki:2013bfa}, 
\begin{align}
 { d\Gamma \over dq^2 } =
 & {G_F^2 |V_{cb}|^2 \over 192\pi^3 m_{B_c}^3} q^2 \sqrt{\lambda_{J/\psi}(q^2)} \left( 1 - {m_\tau^2 \over q^2} \right)^2 \times \notag \\
 & \Bigg\{ ( |1 + C_{V_1}|^2 + |C_{V_2}|^2 ) \left[ \left( 1 + {m_\tau^2 \over2q^2} \right) \left( H_{V_+}^2 + H_{V_-}^2 + H_{V_0}^2 \right) + {3 \over 2}{m_\tau^2 \over q^2} \, H_{V_t}^2 \right] \notag \\
 &~ - 2\text{Re}[(1 + C_{V_1}) C_{V_2}^{*}] \left[ \left( 1 + {m_\tau^2 \over 2q^2} \right) \left( H_{V_0}^2 + 2 H_{V_+} \cdot H_{V_-} \right) + {3 \over 2}{m_\tau^2 \over q^2} \, H_{V_t}^2 \right] \notag \\
 &~ + {3 \over 2} |C_{S_1} - C_{S_2}|^2 \, H_S^2 + 8|C_T|^2 \left( 1+ {2m_\tau^2 \over q^2} \right) \left( H_{T_+}^2 + H_{T_-}^2 + H_{T_0}^2  \right) \notag \\
 &~ + 3\text{Re}[ ( 1 + C_{V_1} - C_{V_2} ) (C_{S_1}^{*} - C_{S_2}^{*} ) ] {m_\tau \over \sqrt{q^2}} \, H_S \cdot H_{V_t} \notag \\
 &~ - 12\text{Re}[ (1 + C_{V_1}) C_T^{*} ] {m_\tau \over \sqrt{q^2}} \left( H_{T_0} \cdot H_{V_0} + H_{T_+} \cdot H_{V_+} - H_{T_-} \cdot H_{V_-} \right) \notag \\
 &~ + 12\text{Re}[ C_{V_2} C_T^{*} ] {m_\tau \over \sqrt{q^2}} \left( H_{T_0} \cdot H_{V_0} + H_{T_+} \cdot H_{V_-} - H_{T_-} \cdot H_{V_+} \right) \Bigg\} \,, 
\end{align}
where $H$s are hadronic helicity amplitudes given by
\begin{align}
  H_{V_\pm}(q^2) 
   & = (m_{B_c}+m_{J/\psi}) A_1^c(q^2) \mp { \sqrt{\lambda_{J/\psi}(q^2)} \over m_{B_c}+m_{J/\psi} } V^c(q^2) \,, \\
  H_{V_0}(q^2) 
   & = { m_{B_c}+m_{J/\psi} \over 2m_{J/\psi}\sqrt{q^2} } \left[ -(m_{B_c}^2-m_{J/\psi}^2-q^2) A_1^c(q^2) + { \lambda_{J/\psi}(q^2) \over (m_{B_c}+m_{J/\psi})^2 } A_2^c(q^2) \right] \,, \\
  H_{V_t}(q^2) 
   & = -\sqrt{ \lambda_{J/\psi}(q^2) \over q^2 } A_0^c(q^2) \,, \\
 \label{EQ:HSeom}
  H_S(q^2) 
   & = -{ \sqrt{\lambda_{J/\psi}(q^2)} \over m_b+m_c } A_0^c(q^2) \,, \\
  H_{T_\pm}(q^2) 
   & = { 1 \over \sqrt{q^2} } \left[ \pm (m_{B_c}^2-m_{J/\psi}^2) T_2^c(q^2) + \sqrt{\lambda_{J/\psi}(q^2)} T_1^c(q^2) \right] \,, \\
   H_{T_0}(q^2) 
    & = { 1 \over 2m_{J/\psi} } \left[ -(m_{B_c}^2+3m_{J/\psi}^2-q^2) T_2^c(q^2) + { \lambda_{J/\psi}(q^2) \over m_{B_c}^2-m_{J/\psi}^2 } T_3^c(q^2) \right] \,,  
\end{align}
and $\lambda_{J/\psi}(q^2) = [ (m_{B_c} - m_{J/\psi})^2 -q^2 ] [ (m_{B_c} + m_{J/\psi})^2 -q^2 ]$. 
The functions $V^c$, $A_i^c$, and $T_i^c$ are form factors (FFs) for the $B_c \to J/\psi$ transition whose definitions are given in Appendix~\ref{FFdef}. 
The scalar hadronic amplitude is obtained as in \eqref{EQ:HSeom} using the quark-level equation of motion.

The FFs for the vector and axial-vector currents have been investigated in Ref.~\cite{Wen-Fei:2013uea} with the use of perturbative QCD~\cite{Beneke:2000wa,Kurimoto:2002sb} and then the following parametrizations are given:  
\begin{align}
 V^c(q^2) & = V^c(0) \exp \Big[ 0.065\, q^2 +0.0015\, (q^2)^2 \Big] \,, \\
 A_0^c(q^2) & = A_0^c(0) \exp \Big[ 0.047\, q^2 +0.0017\, (q^2)^2 \Big] \,, \\
 A_1^c(q^2) & = A_1^c(0) \exp \Big[ 0.038\, q^2 +0.0015\, (q^2)^2 \Big] \,, \\
 A_2^c(q^2) & = A_2^c(0) \exp \Big[ 0.064\, q^2 +0.0041\, (q^2)^2 \Big] \,,  
\end{align}
where the values for the $q^2=0$ point are obtained 
by the fit; $V^c(0) = 0.42 \pm 0.01 \pm 0.01$, $A_0^c(0) = 0.59 \pm 0.02 \pm 0.01$, $A_1^c(0) = 0.46 \pm 0.02 \pm 0.01$, and $A_2^c(0) = 0.64 \pm 0.02 \pm 0.01$~\cite{Wen-Fei:2013uea}. 
As for the tensor FFs, we simply adopt the quark-level equation of motion, (see Ref.~\cite{Sakaki:2013bfa}.) That is, 
\begin{align}
 T_1^c(q^2) & = {m_b+m_c \over m_{B_c} + m_{J/\psi}} V^c(q^2) \,, \\
 T_2^c(q^2) & = {m_b-m_c \over m_{B_c} - m_{J/\psi}} A_1^c(q^2) \,, \\
 T_3^c(q^2) & = -{ m_b-m_c \over q^2 } \Big[ m_{B_c} \big( A_1^c(q^2)-A_2^c(q^2) \big) + m_{J/\psi} \big(A_2^c(q^2)+A_1^c(q^2)-2A_0^c(q^2)\big) \Big] \,. 
\end{align}
Therefore, we are now ready to calculate the decay rate in any type of NP model. 

\section{Numerical analysis}
\label{Sec:analysis}

For numerical evaluation on $R_{J/\psi}$, we take the following values for input; 
$m_{B_c} = 6.275\,\text{GeV}$, $m_{J/\psi} = 3.096\,\text{GeV}$, $m_\tau = 1.777\,\text{GeV}$, $m_b+m_c = 6.2\,\text{GeV}$, and $m_b-m_c = 3.45\,\text{GeV}$~\cite{Patrignani:2016xqp}. 
Then, the SM predicts 
\begin{align}
R_{J/\psi}^\text{SM} = 0.283 \pm 0.048 \,, 
\end{align}
where the uncertainty comes from the inputs of $V^c(0)$, $A_0^c(0)$, $A_1^c(0)$, and $A_2^c(0)$. 
The result is consistent with Refs.~\cite{Ivanov:2005fd,Dutta:2017xmj}. 
This is compared with \eqref{eq:RJpsiLHCb} and thus, 
one finds that there exists a $1.7\sigma$ deviation from the SM, i.e., $[\chi^2]_{J/\psi}^{\text{SM}} \simeq 2.9$. 
Note that the $R_{J/\psi}$ measurement still include a large uncertainty. 
Combined with the $R_{D}$ and $R_{D^*}$ measurements~\cite{Amhis:2016xyh,LHCbStatus}, it turns out $[\chi^2]_{J/\psi+D+D^*}^{\text{SM}} \simeq 22$.

\begin{figure}[t!]
\begin{center}
\includegraphics[viewport=0 0 750 350, width=40em]{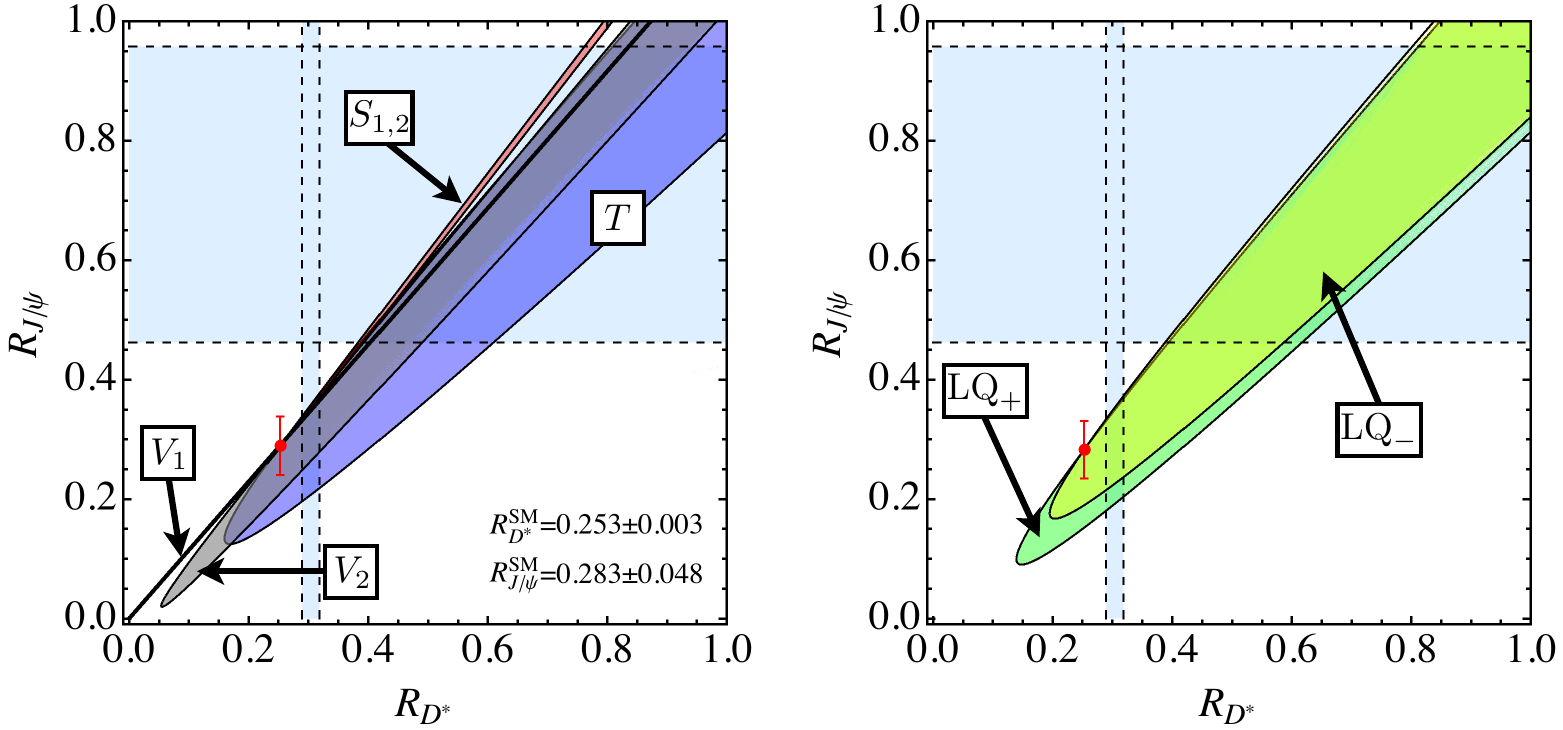}~~~
\caption{
Correlation between $R_{D^*}$ and $R_{J/\psi}$ in the presence of one NP operator, $V_1$, $V_2$, $S_2$, or $T$ (left) and of LQ specific operators with $C_{S_2}=\pm 7.8C_T$ (right). 
The red dot shows the SM predictions with the error bar for $R_{J/\psi}$. 
Note that $S_1$ and $S_2$ have the same contribution. 
}
\label{Fig:RDstRJpsi}
\end{center}
\end{figure}

In fig.\,\ref{Fig:RDstRJpsi}, we show correlation between $R_{D^*}$ and $R_{J/\psi}$ 
in the presence of one NP operator ($V_1$, $V_2$, $S_2$, or $T$) and LQ specific operators ($\text{LQ}_\pm : C_{S_2}=\pm 7.8C_T$), 
where the NP type is denoted in the plot and the dashed lines show the present experimental results at $1\sigma$. 
We see that single NP operators and LQs cannot simultaneously accommodate the present experimental results of $R_{D^*}$ and $R_{J/\psi}$ within $1\sigma$. 
Remind that $C_{V_1} \sim 0.15$ explains the present central values of the $R_{D}$ and $R_{D^*}$ measurements, (for example, see Ref.~\cite{Sakaki:2014sea}.) 
If this is the case, one expects $R_{J/\psi} \sim 0.37$. 
This is one possible way to probe and/or distinguish NP in the $b \to c \tau\nu$ process.

As briefly explained in Sec.\,\ref{Intro}, the lifetime of $B_c$ is a significant tool to constrain NP in $b \to c \tau\bar\nu$, which has been pointed out in Ref.~\cite{Alonso:2016oyd}. 
The idea was as follows. 
The $B_c$ lifetime has been measured as $\tau_{B_c}^\text{exp} = (0.507 \pm 0.008)\,\text{ps}$~\cite{Patrignani:2016xqp}, 
whereas it has been theoretically calculated as $\tau_{B_c}^\text{th} = (0.52^{+0.18}_{-0.12})\,\text{ps}$~\cite{Alonso:2016oyd} within the SM by using an operator product expansion~\cite{Beneke:1996xe}. 
As for the latter, the branching fractions have also been obtained and then 
pure/semi-tauonic modes could have $\lesssim 5\,\%$ for the central value ($\tau_{B_c}^\text{th;cent}$) and $\lesssim 30\,\%$ for the $1\sigma$ upper limit ($\tau_{B_c}^\text{th;$+1\sigma$}$), 
comparing it with $\tau_{B_c}^\text{exp;cent}$. 
Thus, when the branching fraction of $B_c \to \tau\bar\nu$ becomes large as a consequence of explaining the $R_{D^{(*)}}$ anomaly, it would be constrained. 
Since $B_c \to \tau\bar\nu$ is sensitive to the scalar operators $S_{1,2}$, the limit from the $B_c$ lifetime disfavors the possible solution to the $R_{D^{(*)}}$ anomaly by the $S_{1,2}$ operator.

This approach can be further developed by including $B_c \to J/\psi \tau\bar\nu$. 
For simplicity, we take the difference $\delta \Gamma_\text{tot} = 1/\tau_{B_c}^\text{exp;cent} - 1/\tau_{B_c}^\text{th;$+1\sigma$}=0.544 \text{ps}^{-1}$ and demand that the tauonic decay rates do not exceed the difference, namely, 
$\delta \Gamma_\text{tot} > \Gamma (B_c \to \tau\bar\nu) +\Gamma (B_c \to J/\psi \tau\bar\nu)$. 
This condition would give an additional constraint on the NP effect in the $b \to c \tau\nu$ process. 
Recently, Ref.~\cite{Akeroyd:2017mhr} pointed out that LEP1 data taken at the $Z$ peak can give a stronger constraint on the branching fraction of $B_c \to \tau\bar\nu$. 
The conservative limit is then given as $\lesssim 10\,\%$. 
We will also take this limit in the numerical study\footnote{
The way to obtain the limit depends on the theoretical prediction of the branching fraction of $B_c \to J/\psi e\bar\nu$. 
Thus, the NP contribution to $B_c \to J/\psi \tau\bar\nu$ does not affect this result. 
}.  
The branching fraction of $B_c \to \tau\bar\nu$ is written as 
\begin{align}
\mathcal B(B_c \to \tau\bar\nu) 
= 
\tau_{B_c}^\text{exp;cent} {1 \over 8\pi} m_{B_c} m_\tau^2 
\left( 1 - {m_\tau^2 \over m_{B_c}^2} \right)^2 f_{B_c}^2 G_F^2 |V_{cb}|^2 \left| 1+C_{V_1}-C_{V_2} +{m_{B_c}^2 \over m_\tau(m_b+m_c)} (C_{S_1}-C_{S_2}) \right|^2 \,.  
\label{EQ:Bctaunu}
\end{align}
For the analysis, we take $f_{B_c} = 434\,\text{MeV}$ and $|V_{cb}| = 4.09 \times 10^{-2}$. 
We also obtain the favored regions on $C_X$, derived from the $R_D$, $R_{D^*}$, and $R_{J/\psi}$ measurements by simply evaluating $\chi^2$.

\begin{figure}[t!]
\begin{center}
\includegraphics[viewport=0 0 360 362, width=14em]{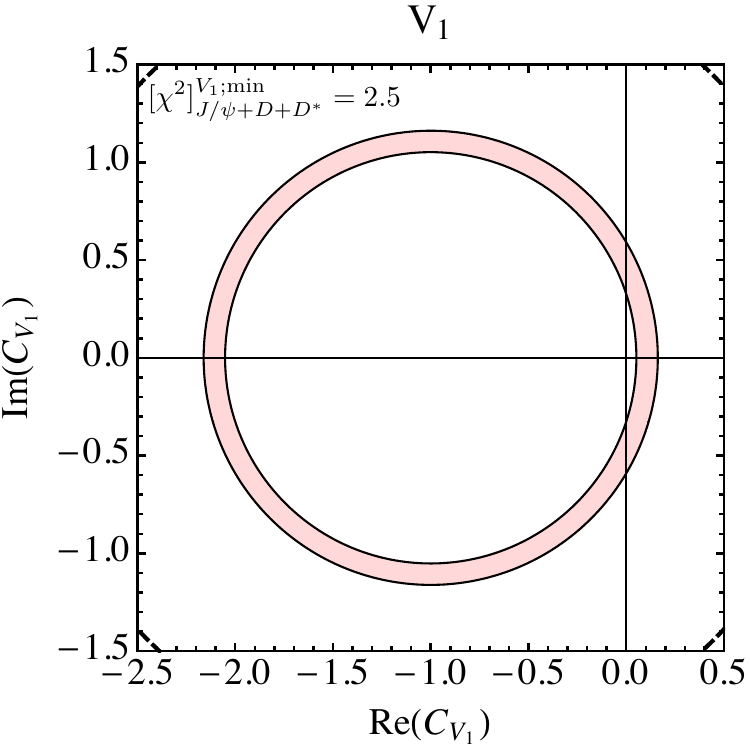}~~
\includegraphics[viewport=0 0 360 362, width=14em]{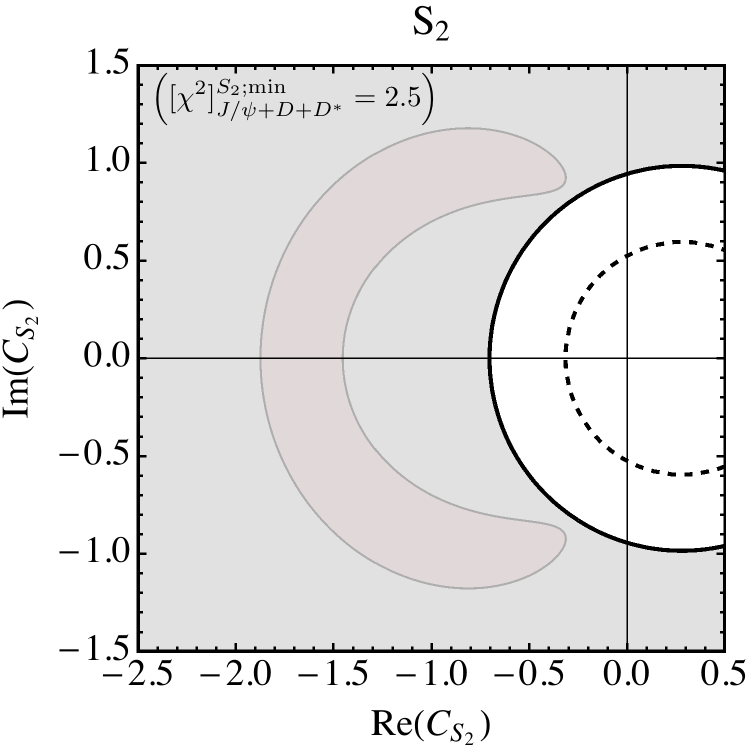}~~
\includegraphics[viewport=0 0 360 362, width=14em]{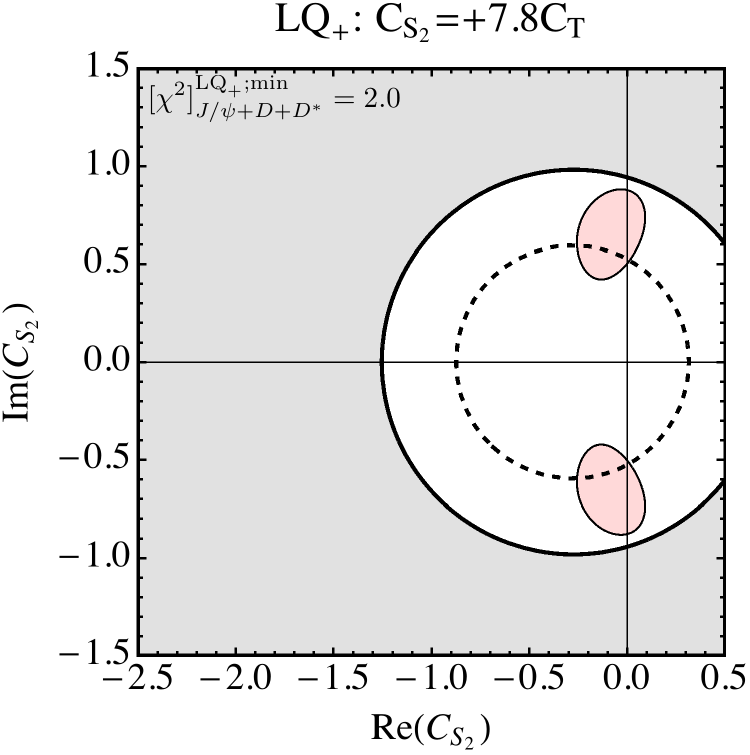} \\[0.5em]
\includegraphics[viewport=0 0 360 362, width=14em]{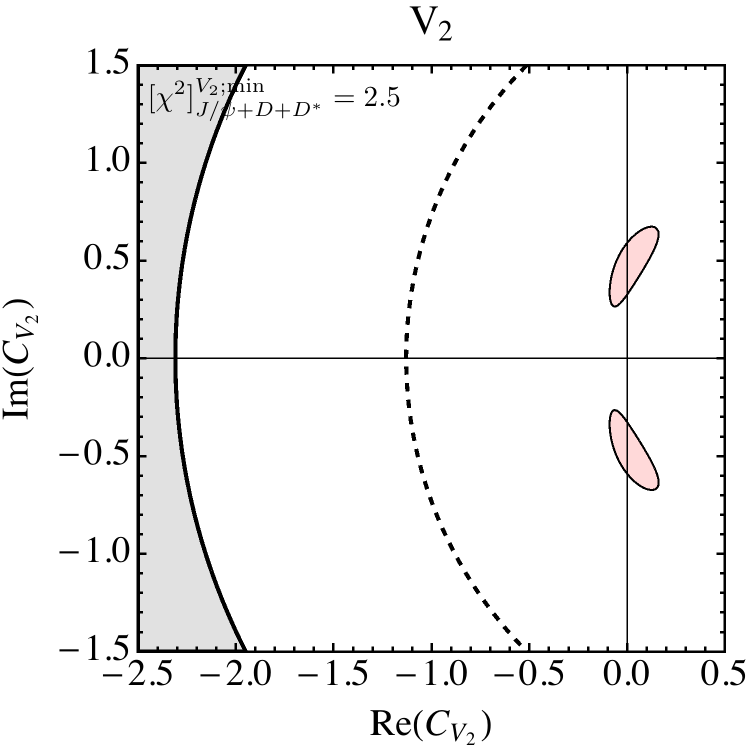}~~
\includegraphics[viewport=0 0 360 362, width=14em]{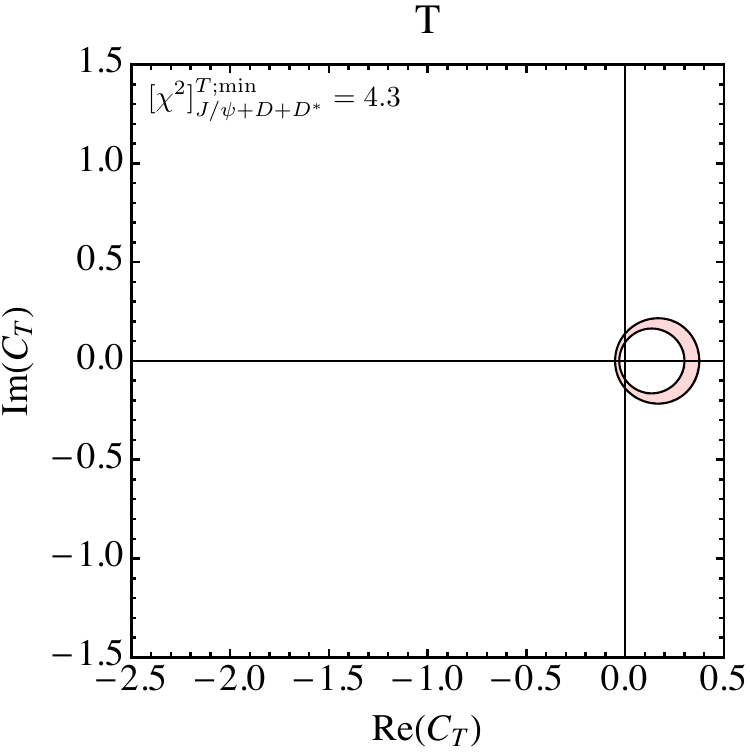}~~
\includegraphics[viewport=0 0 360 362, width=14em]{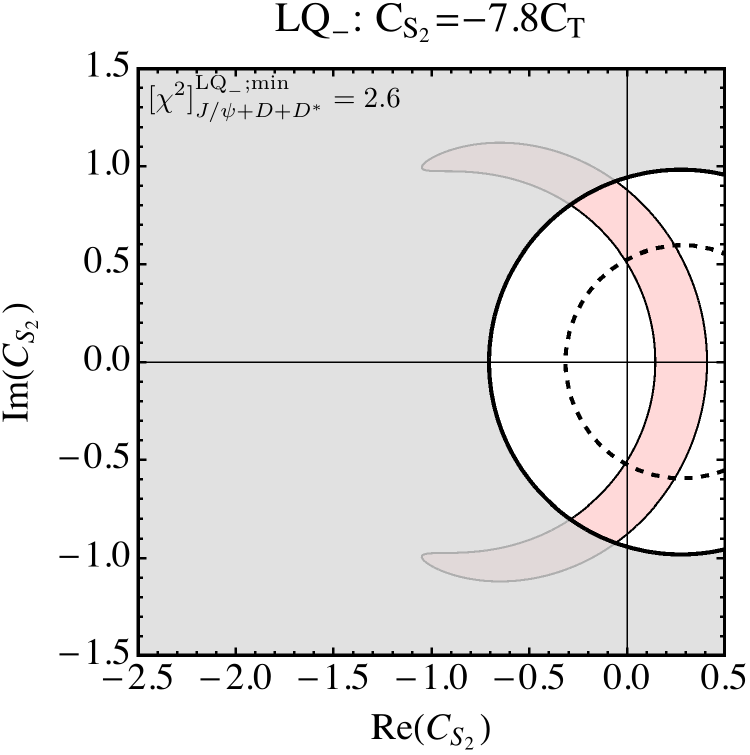}
\caption{
Favored regions from the $R_D$, $R_{D^*}$, and $R_{J/\psi}$ measurements at 95\% CL (red) and 
disfavored regions by the limit from the $B_c$ lifetime (gray with solid boundaries) and from the $B_c\to\tau\bar\nu$ branching ratio (dashed curves), 
in the complex plane of $C_X$ for the $V_1$, $V_2$, $S_1$, $S_2$, $T$, $\text{LQ}_+$, and $\text{LQ}_-$ scenarios. 
A minimal value of $\chi^2$ for each NP scenario is also shown in the legend. 
}
\label{Fig:Allowed}
\end{center}
\end{figure}

In fig.\,\ref{Fig:Allowed}, we show NP bounds in the complex plane of $C_X$ for the $V_1$, $V_2$, $S_2$, $T$, $\text{LQ}_+$, and $\text{LQ}_-$ scenarios. 
Favored regions from the $R_D$, $R_{D^*}$, and $R_{J/\psi}$ measurements, allowed at 95\% confidence level (CL), are shown in red color. 
On the other hand, regions in gray with black solid boundaries are disfavored by the limit from the $B_c$ lifetime, obtained by the aforementioned method. 
The black dashed curves are then the limit obtained by the condition $\mathcal B(B_c \to \tau\bar\nu) \lesssim 10\%$. 
We see that the $S_2$ solution is totally excluded by the $B_c$ lifetime, which is consistent with Ref.~\cite{Alonso:2016oyd}, 
even though the $S_2$ solution has the better fit result to the $R_D$, $R_{D^*}$, and $R_{J/\psi}$ measurements ($[\chi^2]_{J/\psi+D+D^*}^{S_2;\text{min}} \sim 3$) than the SM ($[\chi^2]_{J/\psi+D+D^*}^{\text{SM}} \simeq 22$). 
One also finds that the constraint from the $B_c$ lifetime is significant for the LQ scenarios.  
When we consider the limit on $\mathcal B(B_c \to \tau\bar\nu)$ from the LEP1 data, the $\text{LQ}_+$ solution to the $R_{D^{(*)}}$ anomaly is severely constrained. 
The minimum value of $\chi^2$ for each NP scenario is obtained as exhibited in the plot. 
The $V_1$, $V_2$, $T$, and $\text{LQ}_-$ scenarios have better fit results to the anomaly than the SM, and are consistent with the $B_c$ lifetime and the $\mathcal B(B_c \to \tau\bar\nu)$ limit.

Additional measurements, relevant for $b \to c \tau\nu$, would improve the investigation to probe NP. 
Indeed, the $\tau$ longitudinal polarization in $\bar B \to D^*\tau\bar\nu$ -- 
defined as $P^\tau_{D^*} = {\Gamma^{+}_{D^*} -\Gamma^{-}_{D^*} \over \Gamma^{+}_{D^*} +\Gamma^{-}_{D^*}}$, where $\Gamma^\pm_{D^*}$ is the partial decay rate for the tau helicity to be $\pm 1/2$ --
has been measured by the Belle experiment~\cite{Abdesselam:2016xqt} and thus it would give an additional hint for the NP effect in $b \to c \tau\nu$. 
When we take the best fitted value of $C_X$, obtained from the above study, we can predict $P^\tau_{D^*}$ for each NP scenario. 
The result is shown in Table~\ref{tab:Ptau}. 
We see that the $V_{1,2}$ and $\text{LQ}_\pm$ scenarios predict consistent values with the present Belle result. 
The prediction for the $T$ scenario deviates at $\gtrsim 1\sigma$ from data, which has been pointed out in Ref.~\cite{Alonso:2016oyd}. 
Since the present data includes a large uncertainty, these results are not conclusive yet. 
This study will be improved in the upcoming Belle~II experiment~\cite{Aushev:2010bq}.

\begin{table}[t]
 \begin{center}
 \renewcommand{\arraystretch}{1.5}
 \begin{tabular}{ccccccc}
 \hline\hline
  				&	Belle~\cite{Abdesselam:2016xqt}		&	$V_1$	&	$V_2$	&	$T$			&	$\text{LQ}_+$	&	$\text{LQ}_-$	\\
 \hline
  $P^\tau_{D^*}$ 	&	~~$-0.44 \pm 0.47^{\,+0.20}_{\,-0.17}$~~	&	$-0.50$~	&	~$-0.50$~	&	~$+0.14$~	&	~$-0.41$~		&	~$-0.50$~	\\
 \hline\hline
 \end{tabular}
 \caption{
Predictions on the $\tau$ longitudinal polarization for the best fitted values of $C_X$ which are obtained from the fit to the $R_{D^{(*)}}$ and $R_{J/\psi}$ measurements. 
The present Belle result is also shown. 
 }
 \label{tab:Ptau}
 \end{center}
\end{table}

\section{Summary}
\label{Sec:summary}

We have studied possible NP effects on $B_c \to J/\psi\tau\bar\nu$ in terms of the effective field theory. 
We provided analytic formula for the decay rate of $B_c \to J/\psi\tau\bar\nu$ in the presence of all type of NP operators. 
Given the recently reported data of $R_{J/\psi} = \mathcal B(B_c \to J/\psi \tau\bar\nu)/\mathcal B(B_c \to J/\psi \mu\bar\nu)$ together with the present data of $R_{D^{(*)}}$, 
the discrepancy with the SM prediction reaches $\sim 4.5\sigma$. 
Then it has turned out that the NP scenarios with the $V_1$, $V_2$, $S_2$, $T$, $\text{LQ}_+$, and $\text{LQ}_-$ operators have better fit to the $R_{J/\psi+D+D^*}$ anomaly, 
although a consistent explanation within $1\sigma$ is not available.

On the other hand, the lifetime of $B_c$, considering the NP effect in $B_c \to \tau\bar\nu$ and $B_c \to J/\psi\tau\bar\nu$, gives the useful constraint so that the $S_2$ solution to the $R_{J/\psi+D+D^*}$ anomaly is disfavored. 
The $\text{LQ}_\pm$ solutions are still consistent with, but close to, the limit of the $B_c$ lifetime. 
When we consider the limit on the branching fraction of $B_c \to \tau\bar\nu$ obtained from the LEP1 data, given as $\lesssim 10\%$, the $\text{LQ}_+$ solution is severely constrained. 
The $V_1$, $V_2$, and $T$ solutions are still free from the limits of the $B_c$ lifetime and the $\mathcal B(B_c \to \tau\bar\nu)$.

We have also shown the predictions on the $\tau$ longitudinal polarization obtained by taking the best fit to the $R_{J/\psi+D+D^*}$ measurements for the NP scenarios. 
This is compared with the Belle result and then the predictions for $V_1$, $V_2$, $\text{LQ}_+$, and $\text{LQ}_-$ are still consistent due to the large experimental uncertainty, 
whereas that for $T$ stands at $\gtrsim 1\sigma$.

We expect that these studies will be improved at the Belle~II experiment and by using run 2 data of LHCb. 
In particular, precise measurement of the three ratios ($R_{J/\psi}$, $R_{D}$, $R_{D^*}$) would enable us to test the hypothesis of single operator dominance for NP. 
Further additional measurements regarding the $b \to c \tau\nu$ process, such as $q^2$ distributions of $\bar B \to D^{(*)}\tau\bar\nu$~\cite{Sakaki:2014sea}, definitely give us significant hint for the NP solutions.

\bigskip
\noindent
{\bf Acknowledgments}: 
The author is grateful to David London for a useful discussion on this topic. 
The author thanks Matthias Neubert and Minoru Tanaka for comments on the description of the $B_c \to J/\psi$ transition. 
The author also thanks Andrew Gerard Akeroyd and Chuan-Hung Chen for the comments on the new constraint on the branching fraction of $B_c \to \tau\bar\nu$ by using the LEP1 data given at the $Z$ peak. 
On the present version, the author is grateful to Zhuoran Huang for a comment on a mistake in the numerical calculation for the $S_{1,2}$ case, 
which does not affect the conclusion of this work but does fig.\,\ref{Fig:RDstRJpsi}.

\newpage
\noindent
{\bf Note added}: 
\eqref{EQ:Bctaunu} in the published version was accidentally replaced with the incorrect form. 
To be precise, $\left( 1 - {m_\tau^2 \over m_{B_c}^2} \right)^2$ is correct as written in this paper, although $\left( 1 - {m_\tau^2 \over m_{B_c}^2} \right)$ was placed in the published paper.

\appendix
\section{Form Factors}
\label{FFdef}

Form factors for $B_c \to J/\psi$, given in the literature~\cite{Wen-Fei:2013uea}, are written as 
\begin{align}
  \langle J/\psi | \bar c \gamma^\mu b | B_c \rangle 
  & = {2i V^c(q^2) \over m_{B_c} + m_{J/\psi} }\, \varepsilon^{\mu\nu\rho\sigma}\, \epsilon_\nu^*\, p_\rho^{(B_c)}\, p_\sigma^{(J/\psi)} \,, \\
  \langle J/\psi | \bar c \gamma^\mu\gamma^5 b | B_c \rangle 
  & = 2m_{J/\psi} A_0^c(q^2) { \epsilon^* \cdot q \over q^2 } q^\mu + (m_{B_c} + m_{J/\psi}) A_1^c(q^2) \left[ \epsilon^{*\mu} - { \epsilon^* \cdot q \over q^2 } q^\mu \right] \notag \\
  &~~~ - A_2^c(q^2) { \epsilon^* \cdot q \over m_{B_c} + m_{J/\psi} } \left[ p^{\mu(B_c)} + p^{\mu(J/\psi)} - { m_{B_c}^2 - m_{J/\psi}^2 \over q^2 } q^\mu \right] \,, \\[0.5em]
  \langle J/\psi |\bar c \sigma^{\mu\nu} q_\nu b| B_c \rangle 
  & = 2T_1^c(q^2)\, \varepsilon^{\mu\nu\rho\sigma}  \epsilon_\nu^*\, p_\rho^{(B_c)}\, p_\sigma^{(J/\psi)}  \,, \\[0.5em]
  \langle J/\psi |\bar c \sigma^{\mu\nu}\gamma_5 q_\nu b| B_c \rangle
  & = - T_2^c(q^2) \Big[(m_{B_c}^2-m_{J/\psi}^2) \epsilon^{*\mu} - (\epsilon^* \cdot q) (p^{(B_c)} + p^{(J/\psi)})^\mu\Big] \notag \\
  &~~~ - T_3^c(q^2) (\epsilon^* \cdot q) \left[q^\mu-{q^2 \over m_{B_c}^2-m_{J/\psi}^2} (p^{(B_c)} + p^{(J/\psi)})^\mu \right] \,,  
\end{align}
where the convention $\varepsilon^{0123}=+1$ is taken.

\bibliographystyle{apsrev}

\begin{thebibliography}{10}

\bibitem{Lees:2012xj} 
  J.~P.~Lees {\it et al.} [BaBar Collaboration],
  \href{http://dx.doi.org/10.1103/PhysRevLett.109.101802}{Phys.\ Rev.\ Lett.\  {\bf 109}, 101802 (2012)}
  \href{http://arxiv.org/abs/1205.5442}{{\ttfamily [arXiv:1205.5442 [hep-ex]]}}.

\bibitem{Lees:2013uzd} 
  J.~P.~Lees {\it et al.} [BaBar Collaboration],
  \href{http://dx.doi.org/10.1103/PhysRevD.88.072012}{Phys.\ Rev.\ D {\bf 88}, no. 7, 072012 (2013)} 
  \href{http://arxiv.org/abs/1303.0571}{{\ttfamily  [arXiv:1303.0571 [hep-ex]]}}.

\bibitem{Huschle:2015rga} 
  M.~Huschle {\it et al.} [Belle Collaboration],
  \href{http://dx.doi.org/10.1103/PhysRevD.92.072014}{Phys.\ Rev.\ D {\bf 92}, no. 7, 072014 (2015)}
  \href{http://arxiv.org/abs/1507.03233}{{\ttfamily [arXiv:1507.03233 [hep-ex]]}}.
  
  \bibitem{Sato:2016svk} 
  Y.~Sato {\it et al.} [Belle Collaboration],
  \href{http://dx.doi.org/10.1103/PhysRevD.94.072007}{Phys.\ Rev.\ D {\bf 94}, no. 7, 072007 (2016)}
  \href{http://arxiv.org/abs/1607.07923}{{\ttfamily [arXiv:1607.07923 [hep-ex]]}}.

\bibitem{Abdesselam:2016xqt} 
  A.~Abdesselam {\it et al.},
  \href{http://arxiv.org/abs/1608.06391}{{\ttfamily arXiv:1608.06391 [hep-ex]}}.

\bibitem{Aaij:2015yra} 
  R.~Aaij {\it et al.} [LHCb Collaboration],
  \href{http://dx.doi.org/10.1103/PhysRevLett.115.159901}{Phys.\ Rev.\ Lett.\  {\bf 115}, no. 11, 111803 (2015)} 
  \href{http://dx.doi.org/10.1103/PhysRevLett.115.111803}{Erratum: [Phys.\ Rev.\ Lett.\  {\bf 115}, no. 15, 159901 (2015)]}
  \href{http://arxiv.org/abs/1506.08614}{{\ttfamily [arXiv:1506.08614 [hep-ex]]}}.

\bibitem{Aaij:2017uff} 
  R.~Aaij {\it et al.} [LHCb Collaboration],
  \href{http://arxiv.org/abs/1708.08856}{{\ttfamily arXiv:1708.08856 [hep-ex]}}.


\bibitem{Sakaki:2013bfa} 
  Y.~Sakaki, M.~Tanaka, A.~Tayduganov and R.~Watanabe,
  \href{http://dx.doi.org/10.1103/PhysRevD.88.094012}{Phys.\ Rev.\ D {\bf 88}, no. 9, 094012 (2013)}
  \href{http://arxiv.org/abs/1309.0301}{{\ttfamily [arXiv:1309.0301 [hep-ph]]}}.

\bibitem{Sakaki:2014sea} 
  Y.~Sakaki, M.~Tanaka, A.~Tayduganov and R.~Watanabe,
  \href{http://dx.doi.org/10.1103/PhysRevD.91.114028}{Phys.\ Rev.\ D {\bf 91}, no. 11, 114028 (2015)}
  \href{http://arxiv.org/abs/1412.3761}{{\ttfamily [arXiv:1412.3761 [hep-ph]]}}.

\bibitem{Fajfer:2012vx} 
  S.~Fajfer, J.~F.~Kamenik and I.~Nisandzic,
  \href{http://dx.doi.org/10.1103/PhysRevD.85.094025}{Phys.\ Rev.\ D {\bf 85}, 094025 (2012)}
  \href{http://arxiv.org/abs/1203.2654}{{\ttfamily [arXiv:1203.2654 [hep-ph]]}}.

\bibitem{Becirevic:2012jf} 
  D.~Be\v{c}irevi\'c, N.~Ko\v{s}nik and A.~Tayduganov,
  \href{http://dx.doi.org/10.1016/j.physletb.2012.08.016}{Phys.\ Lett.\ B {\bf 716}, 208 (2012)}
  \href{http://arxiv.org/abs/1206.4977}{{\ttfamily [arXiv:1206.4977 [hep-ph]]}}.

\bibitem{Datta:2012qk} 
  A.~Datta, M.~Duraisamy and D.~Ghosh,
  \href{http://dx.doi.org/10.1103/PhysRevD.86.034027}{Phys.\ Rev.\ D {\bf 86}, 034027 (2012)}
  \href{http://arxiv.org/abs/1206.3760}{{\ttfamily [arXiv:1206.3760 [hep-ph]]}}.

\bibitem{Tanaka:2012nw} 
  M.~Tanaka and R.~Watanabe,
  \href{http://dx.doi.org/10.1103/PhysRevD.87.034028}{Phys.\ Rev.\ D {\bf 87}, no. 3, 034028 (2013)}
  \href{http://arxiv.org/abs/1212.1878}{{\ttfamily [arXiv:1212.1878 [hep-ph]]}}.

\bibitem{Duraisamy:2013kcw} 
  M.~Duraisamy and A.~Datta,
  \href{http://dx.doi.org/10.1007/JHEP09(2013)059}{JHEP {\bf 1309}, 059 (2013)}
  \href{http://arxiv.org/abs/1302.7031}{{\ttfamily [arXiv:1302.7031 [hep-ph]]}}.

\bibitem{Biancofiore:2013ki} 
  P.~Biancofiore, P.~Colangelo and F.~De Fazio,
  \href{http://dx.doi.org/10.1103/PhysRevD.87.074010}{Phys.\ Rev.\ D {\bf 87}, no. 7, 074010 (2013)}
  \href{http://arxiv.org/abs/1302.1042}{{\ttfamily [arXiv:1302.1042 [hep-ph]]}}.

\bibitem{Bhattacharya:2015ida} 
  S.~Bhattacharya, S.~Nandi and S.~K.~Patra,
  \href{http://dx.doi.org/10.1103/PhysRevD.93.034011}{Phys.\ Rev.\ D {\bf 93}, no. 3, 034011 (2016)}
  \href{http://arxiv.org/abs/1509.07259}{{\ttfamily [arXiv:1509.07259 [hep-ph]]}}.

\bibitem{Alok:2016qyh} 
  A.~K.~Alok, D.~Kumar, S.~Kumbhakar and S.~U.~Sankar,
  \href{http://dx.doi.org/10.1103/PhysRevD.95.115038}{Phys.\ Rev.\ D {\bf 95}, no. 11, 115038 (2017)}
  \href{http://arxiv.org/abs/1606.03164}{{\ttfamily [arXiv:1606.03164 [hep-ph]]}}.

\bibitem{Feruglio:2016gvd} 
  F.~Feruglio, P.~Paradisi and A.~Pattori,
  \href{http://dx.doi.org/10.1103/PhysRevLett.118.011801}{Phys.\ Rev.\ Lett.\  {\bf 118}, no. 1, 011801 (2017)}
  \href{http://arxiv.org/abs/1606.00524}{{\ttfamily [arXiv:1606.00524 [hep-ph]]}}.

\bibitem{Ivanov:2016qtw} 
  M.~A.~Ivanov, J.~G.~K\"{o}rner and C.~T.~Tran,
  \href{http://dx.doi.org/10.1103/PhysRevD.94.094028}{Phys.\ Rev.\ D {\bf 94}, no. 9, 094028 (2016)}
  \href{http://arxiv.org/abs/1607.02932}{{\ttfamily [arXiv:1607.02932 [hep-ph]]}}.

\bibitem{Faroughy:2016osc} 
  D.~A.~Faroughy, A.~Greljo and J.~F.~Kamenik,
  \href{http://dx.doi.org/10.1016/j.physletb.2016.11.011}{Phys.\ Lett.\ B {\bf 764}, 126 (2017)}
  \href{http://arxiv.org/abs/1609.07138}{{\ttfamily [arXiv:1609.07138 [hep-ph]]}}.
  
\bibitem{Bardhan:2016uhr} 
  D.~Bardhan, P.~Byakti and D.~Ghosh,
  \href{http://dx.doi.org/10.1007/JHEP01(2017)125}{JHEP {\bf 1701}, 125 (2017)}
  \href{http://arxiv.org/abs/1610.03038}{{\ttfamily [arXiv:1610.03038 [hep-ph]]}}.

\bibitem{Ivanov:2017mrj} 
  M.~A.~Ivanov, J.~G.~K?rner and C.~T.~Tran,
  \href{http://dx.doi.org/10.1103/PhysRevD.95.036021}{Phys.\ Rev.\ D {\bf 95}, no. 3, 036021 (2017)}
  \href{http://arxiv.org/abs/1701.02937}{{\ttfamily [arXiv:1701.02937 [hep-ph]]}}.

\bibitem{Choudhury:2017qyt} 
  D.~Choudhury, A.~Kundu, R.~Mandal and R.~Sinha,
  \href{http://arxiv.org/abs/1706.08437}{{\ttfamily arXiv:1706.08437 [hep-ph]}}.

\bibitem{Hou:1992sy} 
  W.~S.~Hou,
  \href{http://dx.doi.org/10.1103/PhysRevD.48.2342}{Phys.\ Rev.\ D {\bf 48}, 2342 (1993).}

\bibitem{Tanaka:1994ay} 
  M.~Tanaka,
  \href{http://dx.doi.org/10.1007/BF01571294}{Z.\ Phys.\ C {\bf 67}, 321 (1995)}
  \href{http://arxiv.org/abs/hep-ph/9411405}{{\ttfamily   [hep-ph/9411405]}}.

\bibitem{Kiers:1997zt} 
  K.~Kiers and A.~Soni,
  \href{http://dx.doi.org/10.1103/PhysRevD.56.5786}{Phys.\ Rev.\ D {\bf 56}, 5786 (1997)}
  \href{http://arxiv.org/abs/hep-ph/9706337}{{\ttfamily [hep-ph/9706337]}}.
  
\bibitem{Chen:2006nua} 
  C.~H.~Chen and C.~Q.~Geng,
  \href{http://dx.doi.org/10.1088/1126-6708/2006/10/053}{JHEP {\bf 0610}, 053 (2006)}
  \href{http://arxiv.org/abs/hep-ph/0608166}{{\ttfamily [hep-ph/0608166]}}.

\bibitem{Crivellin:2012ye} 
  A.~Crivellin, C.~Greub and A.~Kokulu,
  \href{http://dx.doi.org/10.1103/PhysRevD.86.054014}{Phys.\ Rev.\ D {\bf 86}, 054014 (2012)}
  \href{http://arxiv.org/abs/1206.2634}{{\ttfamily [arXiv:1206.2634 [hep-ph]]}}.

\bibitem{Celis:2012dk} 
  A.~Celis, M.~Jung, X.~Q.~Li and A.~Pich,
  \href{http://dx.doi.org/10.1007/JHEP01(2013)054}{JHEP {\bf 1301}, 054 (2013)}
  \href{http://arxiv.org/abs/1210.8443}{{\ttfamily [arXiv:1210.8443 [hep-ph]]}}.

\bibitem{Crivellin:2013wna} 
  A.~Crivellin, A.~Kokulu and C.~Greub,
  \href{http://dx.doi.org/10.1103/PhysRevD.87.094031}{Phys.\ Rev.\ D {\bf 87}, no. 9, 094031 (2013)}
  \href{http://arxiv.org/abs/1303.5877}{{\ttfamily [arXiv:1303.5877 [hep-ph]]}}.

\bibitem{Cline:2015lqp} 
  J.~M.~Cline,
  \href{http://dx.doi.org/10.1103/PhysRevD.93.075017}{Phys.\ Rev.\ D {\bf 93}, no. 7, 075017 (2016)}
  \href{http://arxiv.org/abs/1512.02210}{{\ttfamily [arXiv:1512.02210 [hep-ph]]}}.

\bibitem{Kim:2015zla} 
  C.~S.~Kim, Y.~W.~Yoon and X.~B.~Yuan,
  \href{http://dx.doi.org/10.1007/JHEP12(2015)038}{JHEP {\bf 1512}, 038 (2015)}
  \href{http://arxiv.org/abs/1509.00491}{{\ttfamily [arXiv:1509.00491 [hep-ph]]}}.
  
\bibitem{Crivellin:2015hha} 
  A.~Crivellin, J.~Heeck and P.~Stoffer,
  \href{http://dx.doi.org/10.1103/PhysRevLett.116.081801}{Phys.\ Rev.\ Lett.\  {\bf 116}, no. 8, 081801 (2016)}
  \href{http://arxiv.org/abs/1507.07567}{{\ttfamily [arXiv:1507.07567 [hep-ph]]}}.

\bibitem{Wang:2016ggf} 
  L.~Wang, J.~M.~Yang and Y.~Zhang,
  \href{http://dx.doi.org/10.1016/j.nuclphysb.2017.09.002}{Nucl.\ Phys.\ B {\bf 924}, 47 (2017)}
  \href{http://arxiv.org/abs/1610.05681}{{\ttfamily [arXiv:1610.05681 [hep-ph]]}}.

\bibitem{Celis:2016azn} 
  A.~Celis, M.~Jung, X.~Q.~Li and A.~Pich,
  \href{http://dx.doi.org/10.1016/j.physletb.2017.05.037}{Phys.\ Lett.\ B {\bf 771}, 168 (2017)}
  \href{http://arxiv.org/abs/1612.07757}{{\ttfamily [1612.07757 [hep-ph]]}}.

\bibitem{Iguro:2017ysu} 
  S.~Iguro and K.~Tobe,
  \href{http://arxiv.org/abs/1708.06176}{{\ttfamily arXiv:1708.06176 [hep-ph]}}.

\bibitem{Ko:2012sv} 
  P.~Ko, Y.~Omura and C.~Yu,
  \href{http://dx.doi.org/10.1007/JHEP03(2013)151}{JHEP {\bf 1303}, 151 (2013)}
  \href{http://arxiv.org/abs/1212.4607}{{\ttfamily [arXiv:1212.4607 [hep-ph]]}}.

\bibitem{Alonso:2015sja} 
  R.~Alonso, B.~Grinstein and J.~Martin Camalich,
  \href{http://dx.doi.org/10.1007/JHEP10(2015)184}{JHEP {\bf 1510}, 184 (2015)}
  \href{http://arxiv.org/abs/1505.05164}{{\ttfamily [arXiv:1505.05164 [hep-ph]]}}.

\bibitem{Freytsis:2015qca} 
  M.~Freytsis, Z.~Ligeti and J.~T.~Ruderman,
  \href{http://dx.doi.org/10.1103/PhysRevD.92.054018}{Phys.\ Rev.\ D {\bf 92}, no. 5, 054018 (2015)}
  \href{http://arxiv.org/abs/1506.08896}{{\ttfamily [arXiv:1506.08896 [hep-ph]]}}.

\bibitem{Barbieri:2015yvd} 
  R.~Barbieri, G.~Isidori, A.~Pattori and F.~Senia,
  \href{http://dx.doi.org/10.1140/epjc/s10052-016-3905-3}{Eur.\ Phys.\ J.\ C {\bf 76}, no. 2, 67 (2016)}
  \href{http://arxiv.org/abs/1512.01560}{{\ttfamily [arXiv:1512.01560 [hep-ph]]}}.

\bibitem{Calibbi:2015kma} 
  L.~Calibbi, A.~Crivellin and T.~Ota,
  \href{http://dx.doi.org/10.1103/PhysRevLett.115.181801}{Phys.\ Rev.\ Lett.\  {\bf 115}, 181801 (2015)}
  \href{http://arxiv.org/abs/1506.02661}{{\ttfamily [arXiv:1506.02661 [hep-ph]]}}.

\bibitem{Fajfer:2015ycq} 
  S.~Fajfer and N.~Ko\v{s}nik,
  \href{http://dx.doi.org/10.1016/j.physletb.2016.02.018}{Phys.\ Lett.\ B {\bf 755}, 270 (2016)}
  \href{http://arxiv.org/abs/1511.06024}{{\ttfamily [arXiv:1511.06024 [hep-ph]]}}.

\bibitem{Bauer:2015knc} 
  M.~Bauer and M.~Neubert,
  \href{http://dx.doi.org/10.1103/PhysRevLett.116.141802}{Phys.\ Rev.\ Lett.\  {\bf 116}, no. 14, 141802 (2016)}
  \href{http://arxiv.org/abs/1511.01900}{{\ttfamily [arXiv:1511.01900 [hep-ph]]}}.

\bibitem{Hati:2015awg} 
  C.~Hati, G.~Kumar and N.~Mahajan,
  \href{http://dx.doi.org/10.1007/JHEP01(2016)117}{JHEP {\bf 1601}, 117 (2016)}
  \href{http://arxiv.org/abs/1511.03290}{{\ttfamily [arXiv:1511.03290 [hep-ph]]}}.
  
\bibitem{Sahoo:2015pzk} 
  S.~Sahoo and R.~Mohanta,
  \href{http://dx.doi.org/10.1103/PhysRevD.93.114001}{Phys.\ Rev.\ D {\bf 93}, no. 11, 114001 (2016)}
  \href{http://arxiv.org/abs/1512.04657}{{\ttfamily [arXiv:1512.04657 [hep-ph]]}}.

\bibitem{Zhu:2016xdg} 
  J.~Zhu, H.~M.~Gan, R.~M.~Wang, Y.~Y.~Fan, Q.~Chang and Y.~G.~Xu,
  \href{http://dx.doi.org/10.1103/PhysRevD.93.094023}{Phys.\ Rev.\ D {\bf 93}, no. 9, 094023 (2016)}
  \href{http://arxiv.org/abs/1602.06491}{{\ttfamily [arXiv:1602.06491 [hep-ph]]}}.
  
\bibitem{Dumont:2016xpj} 
  B.~Dumont, K.~Nishiwaki and R.~Watanabe,
  \href{http://dx.doi.org/10.1103/PhysRevD.94.034001}{Phys.\ Rev.\ D {\bf 94}, no. 3, 034001 (2016)}
  \href{http://arxiv.org/abs/1603.05248}{{\ttfamily [arXiv:1603.05248 [hep-ph]]}}.

\bibitem{Bhattacharya:2014wla} 
  B.~Bhattacharya, A.~Datta, D.~London and S.~Shivashankara,
  \href{http://dx.doi.org/10.1016/j.physletb.2015.02.011}{Phys.\ Lett.\ B {\bf 742}, 370 (2015)}
  \href{http://arxiv.org/abs/1412.7164}{{\ttfamily [arXiv:1412.7164 [hep-ph]]}}.

\bibitem{Das:2016vkr} 
  D.~Das, C.~Hati, G.~Kumar and N.~Mahajan,
  \href{http://dx.doi.org/10.1103/PhysRevD.94.055034}{Phys.\ Rev.\ D {\bf 94}, 055034 (2016)}
  \href{http://arxiv.org/abs/1605.06313}{{\ttfamily [arXiv:1605.06313 [hep-ph]]}}.

\bibitem{Li:2016vvp} 
  X.~Q.~Li, Y.~D.~Yang and X.~Zhang,
  \href{http://dx.doi.org/10.1007/JHEP08(2016)054}{JHEP {\bf 1608}, 054 (2016)}
  \href{http://arxiv.org/abs/1605.09308}{{\ttfamily [arXiv:1605.09308 [hep-ph]]}}.

\bibitem{Bhattacharya:2016mcc} 
  B.~Bhattacharya, A.~Datta, J.~P.~Gu\'evin, D.~London and R.~Watanabe,
  \href{http://dx.doi.org/10.1007/JHEP01(2017)015}{JHEP {\bf 1701}, 015 (2017)}
  \href{http://arxiv.org/abs/1609.09078}{{\ttfamily [arXiv:1609.09078 [hep-ph]]}}.

\bibitem{Becirevic:2016yqi} 
  D.~Be\v{c}irevi\'c, S.~Fajfer, N.~Ko\v{s}nik and O.~Sumensari,
  \href{http://dx.doi.org/10.1103/PhysRevD.94.115021}{Phys.\ Rev.\ D {\bf 94}, no. 11, 115021 (2016)}
  \href{http://arxiv.org/abs/1608.08501}{{\ttfamily [arXiv:1608.08501 [hep-ph]]}}.

\bibitem{Sahoo:2016pet} 
  S.~Sahoo, R.~Mohanta and A.~K.~Giri,
  \href{http://dx.doi.org/10.1103/PhysRevD.95.035027}{Phys.\ Rev.\ D {\bf 95}, no. 3, 035027 (2017)}
  \href{http://arxiv.org/abs/1609.04367}{{\ttfamily [arXiv:1609.04367 [hep-ph]]}}.

\bibitem{Boucenna:2016wpr} 
  S.~M.~Boucenna, A.~Celis, J.~Fuentes-Martin, A.~Vicente and J.~Virto,
  \href{http://dx.doi.org/10.1016/j.physletb.2016.06.067}{Phys.\ Lett.\ B {\bf 760}, 214 (2016)}
  \href{http://arxiv.org/abs/1604.03088}{{\ttfamily [arXiv:1604.03088 [hep-ph]]}}.

\bibitem{Boucenna:2016qad} 
  S.~M.~Boucenna, A.~Celis, J.~Fuentes-Martin, A.~Vicente and J.~Virto,
  \href{http://dx.doi.org/10.1007/JHEP12(2016)059}{JHEP {\bf 1612}, 059 (2016)}
  \href{http://arxiv.org/abs/1608.01349}{{\ttfamily [arXiv:1608.01349 [hep-ph]]}}.

\bibitem{Altmannshofer:2017poe} 
  W.~Altmannshofer, P.~S.~B.~Dev and A.~Soni,
  \href{http://arxiv.org/abs/1704.06659}{{\ttfamily [arXiv:1704.06659 [hep-ph]]}}.

    
\bibitem{Alonso:2016oyd} 
  R.~Alonso, B.~Grinstein and J.~Martin Camalich,
  \href{http://dx.doi.org/10.1103/PhysRevLett.118.081802}{Phys.\ Rev.\ Lett.\  {\bf 118}, no. 8, 081802 (2017)}
  \href{http://arxiv.org/abs/1611.06676}{{\ttfamily [arXiv:1611.06676 [hep-ph]]}}.

\bibitem{Akeroyd:2017mhr} 
  A.~G.~Akeroyd and C.~H.~Chen,
  \href{http://arxiv.org/abs/1708.04072}{{\ttfamily arXiv:1708.04072 [hep-ph]}}.
 
\bibitem{LHCbRJpsi}
  LHCb-PAPER-2017-035. 
  
\bibitem{LHCbStatus}
  Presentation by M.~Fontana, on behalf of LHCb Collaboration, \href{https://indico.cern.ch/event/658856/contributions/2686351/attachments/1522412/2379024/talk_LHCb_MFontana.pdf}{[talk slide]}.
  
\bibitem{Wen-Fei:2013uea} 
  W.~F.~Wang, Y.~Y.~Fan and Z.~J.~Xiao,
  \href{http://dx.doi.org/10.1088/1674-1137/37/9/093102}{Chin.\ Phys.\ C {\bf 37}, 093102 (2013)}
  \href{http://arxiv.org/abs/1212.5903}{{\ttfamily [arXiv:1212.5903 [hep-ph]]}}.
  
\bibitem{Beneke:2000wa} 
  M.~Beneke and T.~Feldmann,
  \href{http://dx.doi.org/10.1016/S0550-3213(00)00585-X}{Nucl.\ Phys.\ B {\bf 592}, 3 (2001)}
  \href{http://arxiv.org/abs/hep-ph/0008255}{{\ttfamily [hep-ph/0008255]}}.  

\bibitem{Kurimoto:2002sb} 
  T.~Kurimoto, H.~n.~Li and A.~I.~Sanda,
  \href{http://dx.doi.org/10.1103/PhysRevD.67.054028}{Phys.\ Rev.\ D {\bf 67}, 054028 (2003)}
  \href{http://arxiv.org/abs/hep-ph/0210289}{{\ttfamily [hep-ph/0210289]}}. 

\bibitem{Patrignani:2016xqp} 
  C.~Patrignani {\it et al.} [Particle Data Group],
  \href{http://dx.doi.org/10.1088/1674-1137/40/10/100001}{Chin.\ Phys.\ C {\bf 40}, no. 10, 100001 (2016).}

\bibitem{Ivanov:2005fd} 
  M.~A.~Ivanov, J.~G.~Korner and P.~Santorelli,
  \href{http://dx.doi.org/10.1103/PhysRevD.75.019901}{Phys.\ Rev.\ D {\bf 71}, 094006 (2005)}
  \href{http://dx.doi.org/10.1103/PhysRevD.71.094006}{Erratum: [Phys.\ Rev.\ D {\bf 75}, 019901 (2007)]}
  \href{http://arxiv.org/abs/hep-ph/0501051}{{\ttfamily [hep-ph/0501051]}}. 

\bibitem{Dutta:2017xmj} 
  R.~Dutta and A.~Bhol,
  \href{http://arxiv.org/abs/1701.08598}{{\ttfamily arXiv:1701.08598 [hep-ph]}}.

\bibitem{Amhis:2016xyh} 
  Y.~Amhis {\it et al.},
  \href{http://arxiv.org/abs/1612.07233}{{\ttfamily arXiv:1612.07233 [hep-ex]}}.

\bibitem{Beneke:1996xe} 
  M.~Beneke and G.~Buchalla,
  \href{http://dx.doi.org/10.1103/PhysRevD.53.4991}{Phys.\ Rev.\ D {\bf 53}, 4991 (1996)}
  \href{http://arxiv.org/abs/hep-ph/9601249}{{\ttfamily [hep-ph/9601249]}}.

\bibitem{Aushev:2010bq} 
  T.~Aushev {\it et al.},
  \href{http://arxiv.org/abs/1002.5012}{{\ttfamily arXiv:1002.5012 [hep-ex]}}.

\end{thebibliography}

\end{document}